\documentclass[a4paper,12pt,final]{amsart}
\pdfoutput=1
\usepackage{geometry}
\usepackage{amsmath}
\usepackage{amsfonts}
\usepackage{amssymb}
\usepackage{amsthm}
\usepackage{dsfont}
\usepackage{color}
\usepackage{epsfig}

\newtheorem*{Thm}{Theorem}

\theoremstyle{definition}

\newtheorem*{Exa}{Example}
\newtheorem*{Rem}{Remark}
\newtheorem*{Ack}{Acknowledgements}
\newcommand{\bC}{\mathbb C}
\newcommand{\bE}{\mathbb E}
\newcommand{\bL}{\mathbb L}
\newcommand{\bN}{\mathbb N}
\newcommand{\bR}{\mathbb R}
\newcommand{\cA}{\mathcal A}

\newcommand{\cE}{\mathcal E}
\newcommand{\cH}{\mathcal H}
\newcommand{\cM}{\mathcal M}
\newcommand{\cP}{\mathcal P}

\newcommand{\her}{^{\operatorname h}}
\newcommand{\ii}{{\operatorname i}}
\newcommand{\id}{\mathds{1}}
\newcommand{\tr}{{\rm tr}}

\newcommand{\clplus}{{\rm cl}^{(+1)}}
\newcommand{\cl}{{\rm cl}^{\rm rI}}
\newcommand{\ext}{{\rm ext}}
\newcommand{\ri}{{\rm ri}}
\def\argmax{\mathop{\text{argmax}}}

\begin{document}
\thispagestyle{empty}
\title[MaxEnt extension]{The MaxEnt extension of a quantum Gibbs family, 
convex geometry and geodesics}
\author{Stephan Weis}
\email{maths@stephan-weis.info}
\address{Max-Planck-Institute for Mathematics in the Sciences, 
Inselstrasse 22, D-04103 Leipzig, Germany}
\date{\today}
\begin{abstract}
We discuss methods to analyze a quantum Gibbs family in the ultra-cold regime 
where the norm closure of the Gibbs family fails due to discontinuities of 
the maximum-entropy inference. The current discussion of maximum-entropy 
inference and irreducible correlation in the area of quantum phase 
transitions is a major motivation for this research. We extend a 
representation of the irreducible correlation from finite temperatures to 
absolute zero.\\[1mm]
\end{abstract}
%
%
%
%
\maketitle
%
%
%
%
\section{Introduction}
\par
Closures of exponential families of probability distributions are 
a point of reference for our analysis of quantum Gibbs families.
The classical setting of finite probability vectors belongs, in the
form of diagonal matrices, to the non-commutative quantum setting of 
density matrices, which we synonymously call states. Weis and Knauf 
\cite{Weis-Knauf} found discontinuities of the maximum-entropy 
inference at the `boundary' of a Gibbs family which consists of 
non-maximal rank states. The classical case is continuous because 
diagonal matrices commute. 
\par
An outstanding motivation for a `boundary' analysis is the heuristic 
by Chen et.\ al \cite{Chen2014} that a discontinuous maximum-entropy 
inference signals quantum phase transitions. Quantum phase transitions
belong to ultra-cold physics, they appear for large inverse temperatures 
near the `boundary' of a Gibbs family.
Our asymptotic theory of Gibbs families supports the calculus of von 
Neumann's maximum-entropy principle in quantum mechanics
\cite{vonNeumann27} which was proposed as an inference method by 
Jaynes~\cite{Jaynes57}.
\par
A novelty concerns irreducible correlation, defined by Linden et al., 
and Zhou \cite{LindenPopescuWootters,Zhou08} using the maximum-entropy 
principle, which can be interpreted as the amount of correlations caused 
by interactions between exactly $k$ bodies, $k\in\bN$. We write this 
function in terms of the divergence 
from a Gibbs family of local Hamiltonians. The maximal rank case for
positive temperatures was done by Zhou \cite{Zhou}, for algorithms see 
Niekamp et al.\ \cite{Niekamp2013}. The classical 
theory was developed by Amari, and Ay et al.\ \cite{Amari,AyOlBeJo}. 
\par
We shed light on a proposal of Liu et al.\ \cite{LiuZengZhou} that 
irreducible correlation signals quantum phase transitions. For this 
we support an idea by Chen et al.\ \cite{Chen2014} by pointing out that 
a discontinuous maximum-entropy inference and a discontinuous 
irreducible correlation are intimately connected. As an example we 
show that the irreducible correlation of three qubits is discontinuous. 
This follows also from the work of Linden et al.\ 
\cite{LindenPopescuWootters}. 
%
%
\section{Closures of statistical models} 
\label{sed:gmle}
\par
We follow Csisz\'ar and Mat\'u\v s' ideas \cite{CsiszarMatus2008}
about defining a maximum likelihood estimate (MLE) when the 
likelihood function has no maximum. This leads to closure concepts.
\par
Let $\mu$ be a non-zero Borel measure on $\bR^n$, $n\in\bN$. The log-Laplace 
transform of $\mu$ is defined for $\vartheta\in\bR^n$ by
$\Lambda(\vartheta):=\ln\int_{\bR^n}e^{\langle\vartheta,x\rangle}\mu(dx)$.
The effective domain of $\Lambda$ is
${\rm dom}(\Lambda):=\{\vartheta\in\bR^n\mid\Lambda(\vartheta)<\infty\}$ and
the corresponding exponential family is
\[\textstyle
\cE
:=\{Q_\vartheta\mid\frac{dQ_\vartheta}{d\mu}
=e^{\langle\vartheta,\,\cdot\,\rangle-\Lambda(\vartheta)}
\hspace{1ex}\mu\hspace{1ex}\mbox{a.\,s.},
\hspace{1ex}\vartheta\in{\rm dom}(\Lambda)\},
\]
where $\frac{dQ_\vartheta}{d\mu}$ is the Radon-Nikodym derivative
of $Q_\vartheta$ with respect to $\mu$. The MLE of the mean $a$ of 
an iid sample from a probability measure $Q_\vartheta$ with unknown 
parameter $\vartheta\in{\rm dom}(\Lambda)$ is defined as a maximizer 
$\vartheta^*$ of the function
\begin{equation}\label{eq:log-density}\textstyle
\vartheta\mapsto\langle\vartheta,a\rangle-\Lambda(\vartheta),
\qquad \vartheta\in{\rm dom}(\Lambda).
\end{equation}
If $a$ is the mean of a distribution $Q_\theta$ 
then the choice $\vartheta^*=\theta$ maximizes (\ref{eq:log-density})
because of
\[\textstyle
[\langle\vartheta^*,a\rangle-\Lambda(\vartheta^*)]
-[\langle\vartheta,a\rangle-\Lambda(\vartheta)]
=D(Q_{\vartheta^*}\|Q_\vartheta),
\qquad \vartheta\in{\rm dom}(\Lambda).
\]
Here $D$ is the Kullback-Leibler divergence which is an asymmetric 
distance, $D(P\|Q):=\int\ln\frac{dP}{dQ}dP$ if $P$ is absolutely 
continuous with respect to $Q$ and otherwise $D(P\|Q):=\infty$.  
\par
It is known \cite{CsiszarMatus2008} that for any $a\in\bR^n$ where 
the supremum in (\ref{eq:log-density}), denoted by $\Psi^*(a)$, is 
finite, there exists a unique probability measure $R^*(a)$ such that
\begin{equation}\label{eq:def-gMLE}\textstyle
\Psi^*(a)-[\langle\vartheta,a\rangle-\Lambda(\vartheta)]
\geq D(R^*(a)\|Q_\vartheta),
\qquad \vartheta\in{\rm dom}(\Lambda).
\end{equation}
The generalized MLE $R^*(a)$ is a natural generalization of the MLE
$\vartheta^*$ and $R^*(a)=Q_{\vartheta^*}$ holds if $\vartheta^*$ exists
\cite{CsiszarMatus2008}.
Equation (\ref{eq:def-gMLE}) shows that $R^*(a)$ lies in 
$\cl(\cE):=\{P\mid \inf_{\vartheta\in{\rm dom}(\Lambda)}D(P\|Q_\vartheta)=0\}$,
called rI-closure of $\cE$ (`rI' reads `reverse I' \cite{CM03}). The total 
variation $\delta(P,Q)$ between two Borel probability measures $P,Q$
is bounded above by the divergence in the Pinsker inequality so 
the total variation closure contains $\cl(\cE)$.
\par
An example of $\cl(\cE)\subsetneq\cl(\cl(\cE))$ for $n=3$ is known \cite{CM04}. 
The analogue of the closure operator $\cl$ for a finite-level quantum system 
belongs to a topology, called rI-topology, and is idempotent \cite{Weis-topo}
but the rI-topology is strictly finer than the norm topology. In the classical 
case (of finite support)  the rI-topology equals the norm topology.  
%
%
\section{Ground state level-crossings}
\par
We discuss limits of Gibbs families, ground state level-crossings of 
Hamiltonians and the heuristics by Chen et al.\ \cite{Chen2014} 
regarding quantum phase transitions.
\par
We consider the matrix algebra $M_d$, $d\in\bN$, of complex $d\times d$ 
matrices with identity $\id$ and, for non-zero projections $p=p^2=p^*\in M_d$,
algebras $\cA=p M_dp=\{pap\mid a\in M_d\}$. The real space of Hamiltonians 
$\cA\her=\{a\in\cA\mid a^*=a\}$ is a Euclidean space with the scalar product 
$\langle a,b\rangle=\tr(ab)$. The Gibbs state of $H\in M_d\her$ at the inverse 
temperature $\beta>0$ is given by $g_H(\beta):=e^{-\beta H}/\tr(e^{-\beta H})$. 
The zero-temperature limit 
\begin{equation}\label{eq:uniform-limit}\textstyle
g^\infty(H):=\lim_{\beta\to+\infty}g_H(\beta)=p/\tr(p)
\end{equation}
is a ground state of $H$. More precisely, $p$ is the projection onto the ground 
state space of $H$, that is the eigenspace of $H$ for the smallest eigenvalue. 
\par
Now we consider a sequence of Hamiltonians $H_i\in M_d\her$, $i=1,\ldots,r$, 
$r\in\bN$. Similarly to the ground state $g^\infty(H)$ of $H$ being a limit 
of the curve $g_H$, the ground states of the Hermitian pencil 
$H(\lambda):=\lambda_1H_1+\cdots+\lambda_rH_r$,
$\lambda=(\lambda_1,\ldots,\lambda_r)\in\bR^r$, are limits of the Gibbs family 
$\cE:=\{e^{H(\lambda)}/\tr(e^{H(\lambda)})\mid\lambda\in\bR^r\}$. Both $\cE$
and the limits shall be studied in terms of expected values. Consider the 
state space $\cM(\cA):=\{\rho\in\cA\mid\rho\succeq 0,\tr(\rho)=1\}$.
The state space $\cM_d:=\cM(M_d)$ of the full algebra represents the physical 
states of the quantum system ($\rho\succeq 0$ means that $\rho$ is positive 
semi-definite). The expected value functional is 
$\bE:M_d\her\to\bR^r$, $a\mapsto(\langle H_1,a\rangle,\ldots,\langle H_r,a\rangle)$.
The convex support $\bL:=\bE(\cM_d)$ consists of all expected values
\cite{Barndorff78,CM03,Weis-supp}.
\par
The von Neumann entropy $H(\rho):=-\tr\,\rho\log(\rho)$ is a measure 
of the uncertainty in $\rho\in\cM_d$ \cite{Jaynes57,Wehrl}. The maximum-entropy 
inference is the map
\[\textstyle
\rho^*:\bL\to\cM_d,\qquad
\alpha\mapsto\argmax\{H(\rho)\mid\rho\in\cM_d,\bE(\rho)=\alpha\}.
\]
According to Jaynes \cite{Jaynes57} the state $\rho^*(\alpha)$ has 
expectation value $\alpha$ and minimal other information.
The Gibbs family $\cE$ contains all states $\rho^*(\alpha)$ of maximal 
rank $d$. Wichmann \cite{Wichmann} has shown that $\rho^*$ 
restricted to the relative interior $\ri(\bL)$ of $\bL$ (interior in the 
affine hull) is a real analytic parametrization of $\cE$, so 
$\rho^*(\ri(\bL))=\cE$ holds.
\par
We discuss ground state limits of $\cE$ and their expected values on the 
relative boundary $\bL\setminus\ri(\bL)$ of $\bL$. For $r=2$ Hamiltonians 
we draw $\bL$ in $x$-$y$-coordinates and we parametrize 
$\tilde{H}(\alpha):=H(\cos(\alpha),\sin(\alpha))$, $\alpha\in\bR$. 
We use the Pauli matrices
\[\textstyle
\sigma_1:=\left[\begin{array}{cc} 0 & 1 \\ 1 & 0 \end{array}\right],\quad
\sigma_2:=\left[\begin{array}{cc} 0 & -\ii \\ \ii & 0 \end{array}\right],\quad
\sigma_3:=\left[\begin{array}{cc} 1 & 0 \\0 & -1 \end{array}\right]. 
\]
We denote the standard basis of $\bC^3$ by $e_1,e_2,e_3$ and we 
write $v^*w$ for the inner product of $v,w\in\bC^3$ and $vw^*$ 
for the linear map $vw^*(z):=(w^*z)\cdot v$ defined for $z\in\bC^3$.
\begin{figure}[t]
a) \includegraphics[height=3cm]{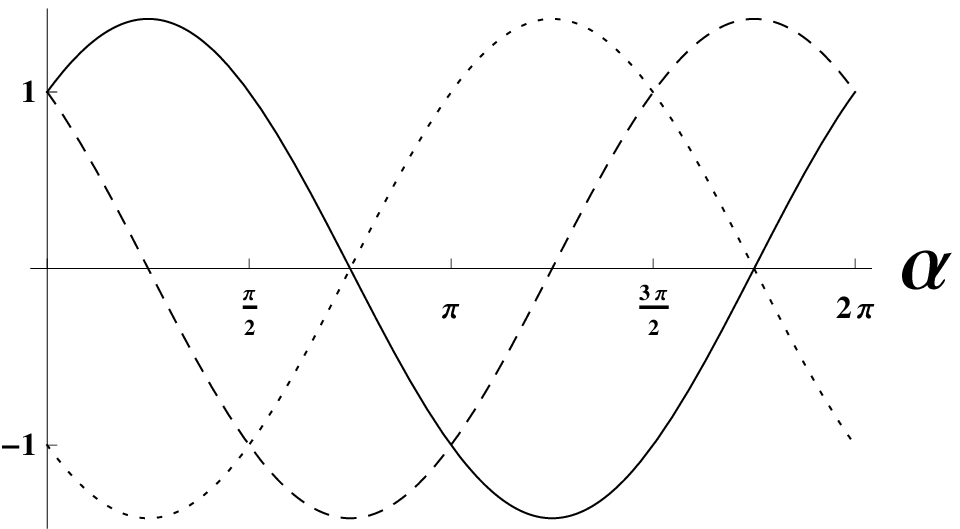}\qquad
b) \includegraphics[height=4cm]{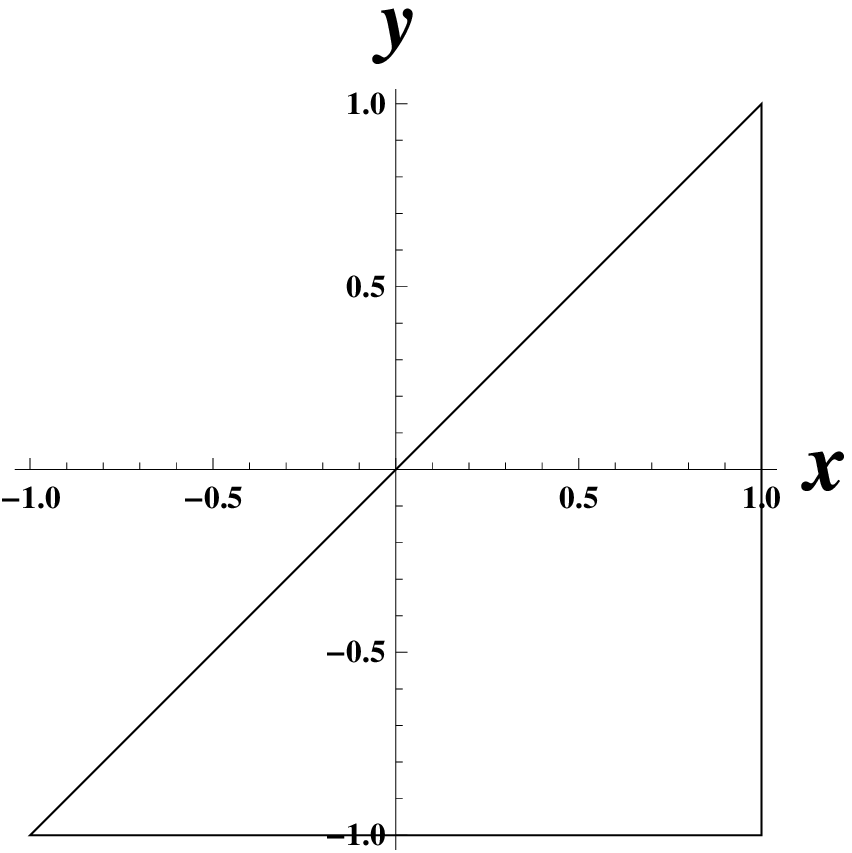}
\caption{\label{fig1}
a) Eigenvalues of $\tilde{H}$ with Type I level-crossing 
b) Triangle convex support $\bL$.}
\end{figure}
\par
The first example of a level-crossing has a discontinuous ground state 
with discontinuous expected value and is called Type I level-crossing by 
Chen et al.\ \cite{Chen2014}. It occurs in the commutative case of finite 
systems and corresponds to a first-order phase transition. 
\begin{Exa}[Type I level-crossing]
Let $H_1=\sigma_3\oplus 1$ and $H_2=\sigma_3\oplus(-1)$, the direct sums 
being embedded as block diagonal matrices into $M_3$. 
The ground state $g^\infty(\tilde{H}(\alpha))=p/\tr(p)$
stays at $p=e_2e_2^*$ for $-\tfrac{1}{4}\pi<\alpha<\tfrac{1}{2}\pi$ and
jumps to $(e_2e_2^*+e_3e_3^*)/2$ at $\alpha=\tfrac{1}{2}\pi$.
Thereby the expected value $\bE(g^\infty(\tilde{H}(\alpha)))$ jumps from 
$(-1,-1)$ to $(0,-1)$. See Figure~\ref{fig1}.  
\end{Exa}
\begin{figure}[t]
a) \includegraphics[height=3cm]{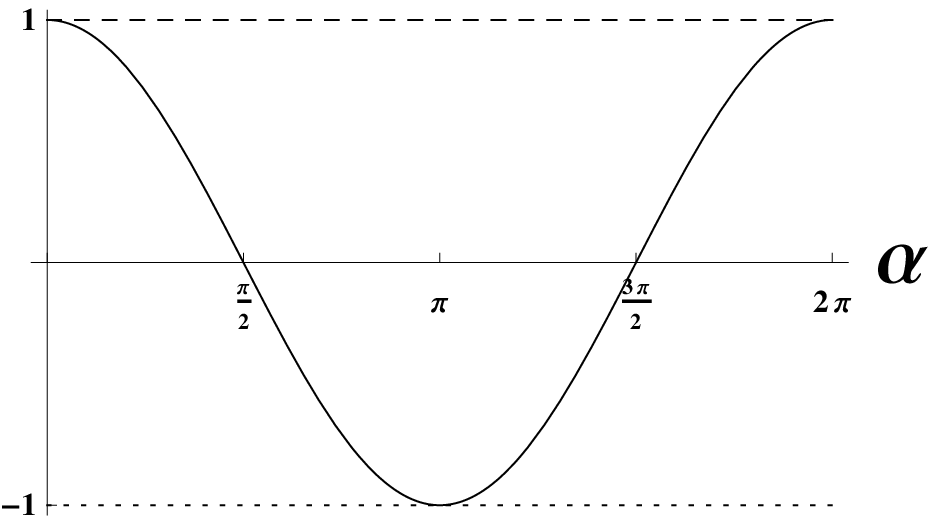}\qquad
b) \includegraphics[height=4cm]{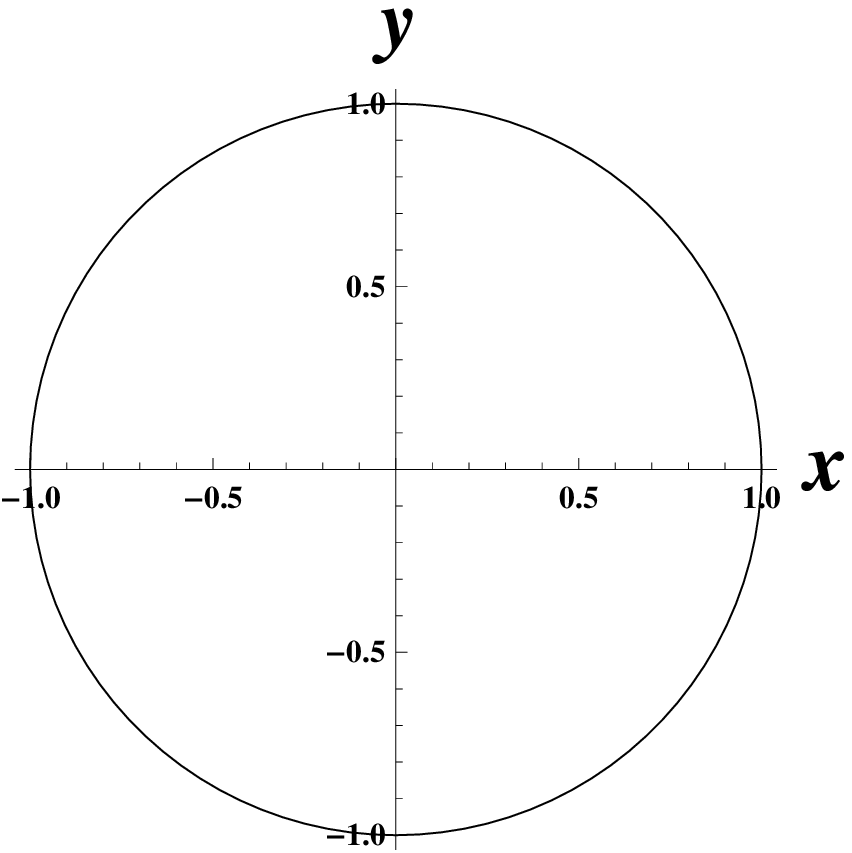}
\caption{\label{fig2}
a) Eigenvalues of $\tilde{H}$ with Type II level-crossing 
b) Disk convex support $\bL$.}
\end{figure}
\par
The second example of a level-crossing has a discontinuous ground state 
with continuous expected value and is called Type II level-crossing 
\cite{Chen2014}. We will see it implies a discontinuous $\rho^*$. 
In the thermodynamic limit a Type II level-crossing is associated with 
a continuous phase transition which includes many quantum phase 
transitions. 
\begin{Exa}[Type II level-crossing]
Let $H_1=\sigma_1\oplus 1$ and $H_2=\sigma_2\oplus0$. The ground state 
is $g^\infty(\tilde{H}(\alpha))=v(\alpha)v(\alpha)^*$ for 
$\alpha\neq\pi\,{\rm mod}\,2\pi$ where $v(\alpha)=(1,-e^{\ii\alpha},0)/\sqrt{2}$ and
$g^\infty(\tilde{H}(\pi))=(v(\pi)v(\pi)^*+e_3e_3^*)/2$. The expected 
value is $\bE(g^\infty(\tilde{H}(\alpha)))=-(\cos(\alpha),\sin(\alpha))$ 
for all $\alpha$. See Figure~\ref{fig2}. 
\end{Exa}
\par 
The last example, a Type I level-crossing, demonstrates a convex geometric
feature.
\begin{figure}[t]
a) \includegraphics[height=3cm]{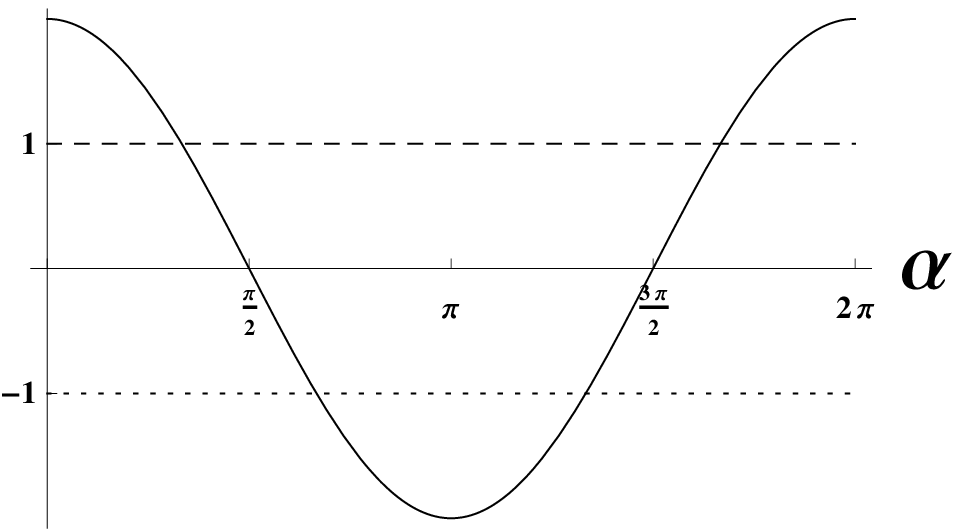}\qquad
b) \includegraphics[height=4cm]{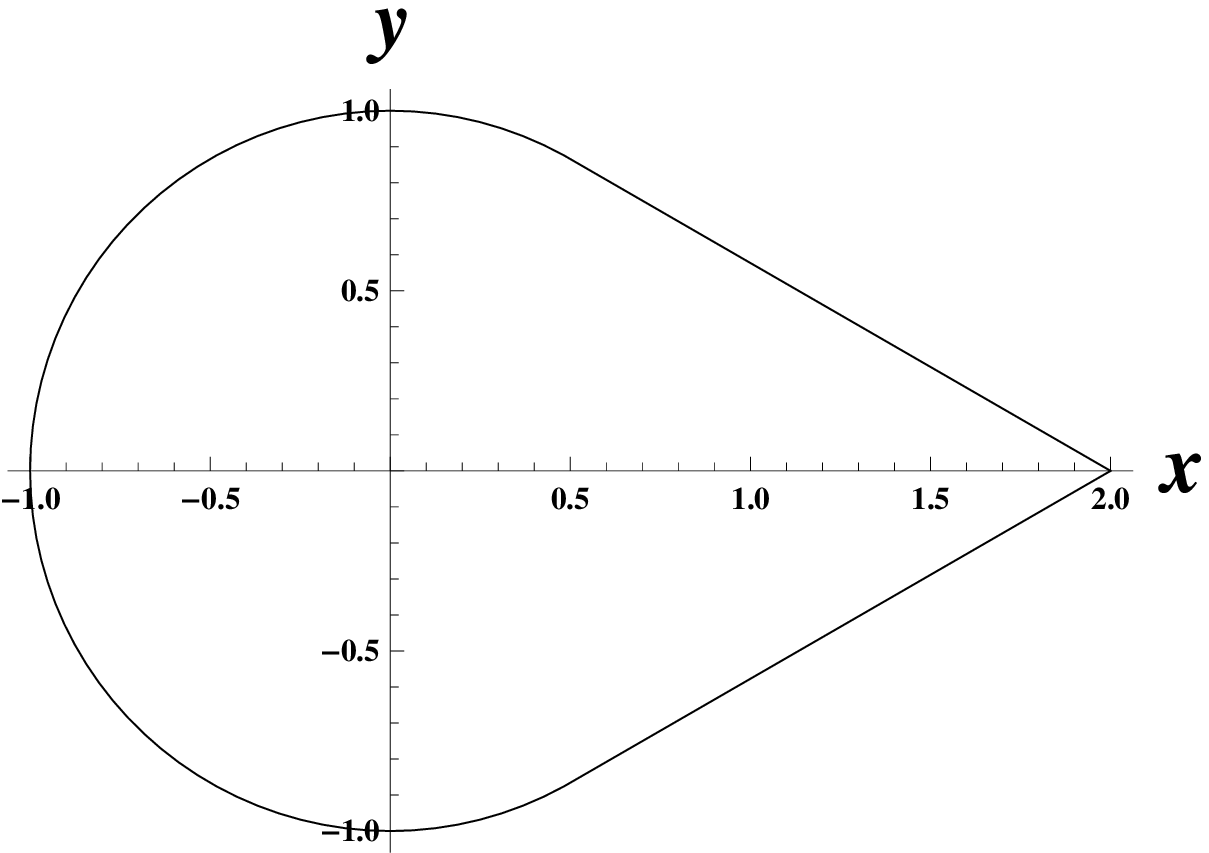}
\caption{\label{fig3}
a) Eigenvalues of $\tilde{H}$ with Type I level-crossing 
b) Drop shape convex support 
$\bL$ (convex hull of the unit disk and the point $(2,0)$) with two 
non-exposed points where the boundary segments meet the disk.}
\end{figure}
\begin{Exa}[Drop shape convex support]
Let $H_1=\sigma_1\oplus 2$ and $H_2=\sigma_2\oplus0$. Ground states
are discussed in Sec.~33(3) in \cite{Weis-supp}. See Figure~\ref{fig3}.
\end{Exa}
%
%
\section{The MaxEnt extension}
\par
We construct an extension of a Gibbs family $\cE$ for which a Pythagorean 
theorem holds. As a corollary the extension is the set of maximum-entropy 
states. 
\par
Maximum-entropy states are unique so there is only one extension with 
Pythagorean theorem. We begin with a geodesic closure 
defined by adding states with maximal support on the ground state space 
of some $H(\lambda)$. Non-exposed points of $\bL$ force us to include 
states without maximal support on the ground state space of any $H(\lambda)$.
\par
The six limit points of the form (\ref{eq:uniform-limit}) in the 
first example above define six expected values. They do not 
cover the relative boundary of the triangle $\bL$ so Wichmann's equation
$\bE(\cE)=\ri(\bL)$ shows that a larger class of curves is needed. A 
$(+1)$-geodesic \cite{Amari-Nagaoka} in the manifold of invertible 
states in $\cM_d$ is defined for $H_0,H\in M_d\her$ by
\[\textstyle
g_{H_0,H}(t):=e^{H_0+tH}/\tr(e^{H_0+tH}),
\qquad t\in\bR.
\]
If $p$ is the projection onto the ground state space of $-H$ then 
Lemma~6.13 in \cite{Weis-topo} shows
\begin{equation}\label{eq:limit}\textstyle
\lim_{t\to\infty}g_{H_0,H}(t)=pe^{pH_0p}/\tr(pe^{pH_0p}).
\end{equation}
The $(+1)$-geodesic closure $\clplus(\cE)$ of the Gibbs family $\cE$ is 
defined as the union of $(+1)$-geodesics in $\cE$ with their limit points.
Let $p\in M_d$ be a non-zero projection and put
$\cE_p:=\left\{pe^{pH(\lambda)p}/\tr(pe^{pH(\lambda)p})\mid\lambda\in\bR^r\right\}$.
The limit (\ref{eq:limit}) shows $\clplus(\cE)=\bigcup_p\cE_p$
where the union is over projections $p$ onto a ground state space of some
$H(\lambda)$, $\lambda\in\bR^r$. 
\par
Convex geometry prevents the inclusion $\bE(\clplus(\cE))\subset\bL$ from
always being an equality, a detailed discussion is Sec.~6.6 in \cite{Weis-topo}. 
If $C\neq\emptyset$ is a compact convex subset of a Euclidean space $X$ then 
$u\in X$ defines an exposed face 
$F_C(u):=\argmax\{\langle c,u\rangle\mid c\in C\}$ of $C$. Exposed faces of 
$\cM_d$ have the form
\begin{equation}\textstyle\label{eq:exp-state}
F_{\cM_d}(H)=\cM(pM_dp),
\qquad H\in M_d\her,
\end{equation}
where $p$ is the projection onto the ground state space of $-H$ 
\cite{Weis-supp}. Exposed faces of the convex support satisfy
\begin{equation}\textstyle\label{eq:exp-cs} 
\bE|_{\cM_d}^{-1}\big(F_\bL(\lambda)\big)
=F_{\cM_d}(H(\lambda)),
\qquad \lambda\in\bR^r.
\end{equation}
For example in Fig.~\ref{fig1}b) we have for 
$\alpha\in(\tfrac{3}{4}\pi,\tfrac{3}{2}\pi)$ and 
$\lambda=(\cos(\alpha),\sin(\alpha))$ exposed faces
$F_\bL(\lambda)=\{(-1,-1)\}$ and $F_{\cM_3}(H(\lambda))=\{e_2e_2^*\}$. 
\par
Wichmann's equality $\rho^*(\ri(\bL))=\cE$ with $\bL$ replaced by an 
exposed face $F$ of $\bL$ and with $\cE$ replaced by $\cE_p$, where 
$p=p(F)$ is defined in (\ref{eq:exp-state}), (\ref{eq:exp-cs}), implies 
that $\bE(\clplus(\cE))$ covers only points of $\bL$ which belong to the 
relative interior of an exposed face of $\bL$. Points not having this 
form, see Fig.~\ref{fig3}b), are called non-exposed points. 
\par
Non-exposed points---and higher-dimensional analogues---have to 
be treated separately, see Sec.~6.2 in \cite{Weis-topo} for details. 
A face of a compact 
convex subset $C$ of a Euclidean space is a convex subset $F$ of $C$ such 
that every segment in $C$ which meets with its relative interior the set $F$ 
belongs 
to $F$. Let $\cP$ denote the set of projections $p$ defined implicitly 
by $\cM(pM_dp)=\bE|_{\cM_d}^{-1}(F)$ for non-empty faces $F$ of $\bL$. Then 
the extension
\[\textstyle
\ext(\cE):=\bigcup_{p\in\cP}\cE_p
\]
induces a bijection $\bE|_{\ext(\cE)}:\ext(\cE)\to\bL$ and we have
$\cE\subset\clplus(\cE)\subset\ext(\cE)$. For $\rho\in\cM_d$ 
we denote by $\pi_\cE(\rho)$ the unique state in $\ext(\cE)$ such that 
$\bE(\rho)=\bE(\pi_\cE(\rho))$. So $\pi_\cE$ is a projection from $\cM_d$ 
onto $\ext(\cE)$.
\par
Pythagorean and projection theorems will show that $\ext(\cE)$ is a useful 
extension. The relative entropy of $\rho,\sigma\in\cM_d$, also known as 
divergence, is defined by $D(\rho,\sigma):=\tr\,\rho(\log(\rho)-\log(\sigma))$ if 
${\rm ker}(\sigma)\subset{\rm ker}(\rho)$. Otherwise $D(\rho,\sigma)=+\infty$. 
The divergence, an asymmetric distance, is non-negative and zero only for 
identical arguments \cite{Wehrl}.
\begin{Thm}[Pythagorean theorem, Thm.~6.12 in \cite{Weis-topo}]
Let $\rho\in\cM_d$ and $\sigma\in\ext(\cE)$. Then 
$D(\rho,\sigma)=D(\rho,\pi_\cE(\rho))+D(\pi_\cE(\rho),\sigma)$
holds.
\end{Thm}
\par
This theorem extends results by Petz, and Amari and Nagaoka 
\cite{Petz94,Amari-Nagaoka} to non-maximal rank states and classical 
results by Csisz\'ar and Mat\'u\v s \cite{CM03} to quantum states. 
\par
The Pythagorean theorem with $\sigma=\id/d$ shows
$\pi_\cE(\rho^*(\alpha))=\rho^*(\alpha)$ for $\alpha\in\bL$, that is 
$\rho^*(\bL)=\ext(\cE)$ holds (details in Sec.~3.4 in \cite{Weis-topo}). In 
the Type II level-crossing example the ground states 
$g^\infty(\tilde{H}(\alpha))$ belong to $\clplus(\cE)\subset\rho^*(\bL)$.
As $g^\infty(\tilde{H}(\alpha))$ is discontinuous at 
$\alpha=\pi$ and has continuous expected values, $\rho^*$ is 
discontinuous at $\bE(g^\infty(\tilde{H}(\pi)))=(1,0)$. This proof is 
similar to Exa.~1 in \cite{Chen2014}, see 
\cite{Weis-Knauf,Weis-cont} for other proofs.
%
%
\section{The reverse I-closure}
\par
A projection theorem allows us to represent the divergence from a Gibbs 
family as a difference of von Neumann entropies. This applies to the 
irreducible correlation.
\par
The divergence from a subset $X\subset\cM_d$ is 
${\rm d}_X:\cM_d\to[0,\infty]$,
$\rho\mapsto\inf\{D(\rho,\sigma)\mid\sigma\in X\}$. Since $\id/d\in\cE$ 
holds, the divergence ${\rm d}_\cE$ has on $\cM_d$ the global upper 
bound $\log(d)$. 
\begin{Thm}[Projection theorem, Thm.~6.16 in \cite{Weis-topo}] 
Let $\rho\in\cM_d$. Then $D(\rho,\,\cdot\,)$ has on $\ext(\cE)$ a unique 
local and global minimum at $\pi_\cE(\rho)$ and 
${\rm d}_\cE(\rho)=D(\rho,\pi_\cE(\rho))$ holds.
\end{Thm}
\par
This theorem shows that $\ext(\cE)$ is the rI-closure 
$\{\rho\in\cM_d\mid{\rm d}_\cE(\rho)=0\}$ of $\cE$. The proof of the
theorem needs Gr\"unbaum's \cite{Gruenbaum} notion of poonem of $\bL$
(Sec.\ 3.6 in \cite{Weis-topo}) which is equivalent to face of $\bL$
and to access sequence \cite{CM05}. Recursively defined, $\bL$ is a 
poonem of $\bL$ and all exposed faces of poonems of $\bL$ are poonems
of $\bL$. In Figure~\ref{fig3}b) the non-exposed points are poonems 
because they are exposed faces of a segment. 
\par
Consider the Gibbs family $\cE(\cH):=\{e^H/\tr(e^H)\mid H\in\cH\}$ of 
a subspace $\cH\subset M_d\her$. If a basis of $\cH$ is chosen then
expectation $\bE$, convex support $\bL$, maximum-entropy inference 
$\rho^*$ and projection $\pi_{\cE(\cH)}$ are defined ($\bE$, $\bL$ and 
$\rho^*$ do not depend much on the basis and $\pi_{\cE(\cH)}$ is 
invariant under basis change). The projection theorem and the 
Pythagorean theorem (with $\sigma=\id/d$) show for $\rho\in\cM_d$ the 
equality
\begin{align}\label{eq:de-diff}\textstyle
{\rm d}_{\cE(\cH)}(\rho)
=D(\rho,\pi_{\cE(\cH)}(\rho))
&=D(\rho,\id/d)-D(\pi_{\cE(\cH)}(\rho),\id/d)\\\nonumber
&=H(\pi_{\cE(\cH)}(\rho))-H(\rho).
\end{align}
Now we consider a flag $\cH_1\subset\cdots\subset\cH_N\subset M_d\her$,
$N\in\bN$, and 
$C_k(\rho):=H(\pi_{\cE(\cH_{k-1})}(\rho))-H(\pi_{\cE(\cH_k)}(\rho))$,
$k=2,\ldots,N$. 
We obtain $C_k={\rm d}_{\cE(\cH_{k-1})}-{\rm d}_{\cE(\cH_k)}$ from 
(\ref{eq:de-diff}). Often $\cH_N=\cA\her$ holds. Then 
${\rm d}_{\cE(\cH_N)}\equiv 0$ follows and we have
${\rm d}_{\cE(\cH_1)}=C_2+\cdots+C_N$.
\par
We turn to the irreducible $k$-body correlation, $k=2,\ldots,N$, of 
a quantum system composed of $N\in\bN$ units. Let the unit 
$i\in[N]:=\{1,\ldots,N\}$ have algebra $\cA_i$ and the total system 
have the tensor product algebra $\cA:=\bigotimes_{i\in[N]}\cA_i$. For 
$A\subset[N]$ let $\cA_A:=\bigotimes_{i\in A}\cA_i$ and let $\id_A$ be
the identity in $\cA_A$. A $k$-local Hamiltonian $H$ is 
a sum of terms of the form $\id_{[N]\setminus A}\otimes b$ for 
$b\in\cA_A\her$ where $A\subset[N]$ has cardinality $|A|\leq k$. 
\par
Let $\cH_k$ be the space of $k$-local Hamiltonians. Then $C_k$ is the 
$k$-body irreducible correlation \cite{LindenPopescuWootters,Zhou08}. 
If  $A\subset[N]$ and $\rho\in\cM(\cA)$ then 
$\langle X,\rho_A\rangle=\langle X\otimes\id_{[N]\setminus A},\rho\rangle$,
$X\in\cA_A\her$, defines a marginal $\rho_A$. The expectation 
$\langle \rho,H\rangle$ of a $k$-local Hamiltonian $H$ can be computed 
from the $k$-reduced density matrices 
$\rho^{(k)}=(\rho_A)_{A\subset[N],|A|=k}$ ($k$-RDM's). So 
$\pi_{\cE(\cH_k)}(\rho)=\rho^*(\rho^{(k)})$ is well-defined and, according 
to Jaynes \cite{Jaynes57}, $\rho^*(\rho^{(k)})$ represents the $k$-RDM's 
of $\rho$ in the most unbiased way because it has minimal other 
information. Zhou \cite{Zhou08} has interpreted $C_k(\rho)$ as the amount 
of $k$-body correlations in $\rho$ which are no $(k-1)$-body correlations 
by arguing that correlation decreases uncertainty. The total correlation 
$I(\rho):=\sum_{i\in[N]}H(\rho_{\{i\}})-H(\rho)$ is also known as 
multi-information \cite{AyKnauf}. 
\par
Since $I={\rm d}_{\cE(\cH_1)}$ holds \cite{Weis-hier} we have, 
as in the paragraph of (\ref{eq:de-diff}), for $k=2,\ldots,N-1$
\begin{equation}\textstyle
C_k={\rm d}_{\cE(\cH_{k-1})}-{\rm d}_{\cE(\cH_k)},
\quad
C_N={\rm d}_{\cE(\cH_{N-1})} \quad \mbox{and} \quad
I=C_2+\cdots+C_N.
\end{equation}
The projection theorem shows, continuing Jaynes' view, that 
${\rm d}_{\cE(\cH_{k-1})}(\rho)$ is the divergence from the 
set of most unbiased representatives of $(k-1)$-RDM's. So it 
is reasonable to interpret it as the amount of correlations 
in $\rho$ caused by interactions of $k$ or more bodies. Then 
$C_k$ is the amount of correlations in $\rho$ caused by 
interactions of exactly $k$ bodies. No information-theoretic 
proof exists for this interpretation except for the total 
correlation \cite{Groisman}.
\par
We point out that $C_3={\rm d}_{\cE(\cH_2)}$ is discontinuous for 
three qubits. Exa.~6 in \cite{Chen2014} shows
that $\rho^*(\rho^{(2)})$ is discontinuous at some 
$\rho^{(2)}=(\rho_{\{2,3\}},\rho_{\{1,3\}},\rho_{\{1,2\}})$ so
Lemma~5.14(2) in \cite{Weis-cont} shows that ${\rm d}_{\cE(\cH_2)}$ 
is discontinuous at some $\sigma\in\cM(\cA)$ with
$\sigma^{(2)}=\rho^{(2)}$. While Lemma~5.14(2) in \cite{Weis-cont}
applies to any composite system and to $\rho^*(\rho^{(k)})$ for any $k$, 
the three qubit discontinuity follows also from the fact
\cite{LindenPopescuWootters} that $C_3(\rho)=0$ holds for pure states 
$\rho$ which are not local 
unitary equivalent to $a|000\rangle+b|111\rangle$. Classically, 
${\rm d}_\cE$ is continuous for any Gibbs family $\cE$, see Sec.\ 6.6 
in \cite{Weis-topo}, so $C_k$ is continuous for all $k$.
%
%
\section{$(-1)$-geodesics}
\par
We now consider geodesics in the Gibbs family $\cE$ which, unlike the 
$(+1)$-geodesics, generate the set of maximum-entropy states $\rho^*(\bL)$ 
with their limits.
\par
A topological analysis \cite{Weis-cont}, related to multi-valued maps 
\cite{Corey}, shows the following.
\begin{Rem}[Polytopes, Thm.~4.9 and Coro.~4.13 in 
\cite{Weis-cont}]
If $X\subset\bL$ is a polytope then $\rho^*|_X$ is continuous. 
So, if $s\subset\ri(\bL)$ is a segment with norm 
closure $\overline{s}$ then $\rho^*|_{\overline{s}}$ is 
continuous.
\end{Rem}
\par
An unparametrized $(-1)$-geodesic in $\cE$ is defined as the 
image $\rho^*(s)$ of a relative open segment 
$s\subset\ri(\bL)$, see \cite{Amari-Nagaoka}. This definition is
consistent by Wichmann's equation $\rho^*(\ri(\bL))=\cE$. Since 
$\rho^*|_{\overline{s}}$ is continuous we get the following.
\begin{Thm}[Geodesic closure, Thm.~5.10 in \cite{Weis-cont}]
The union of $(-1)$-geodesics in $\cE$ and their limit points
equals $\rho^*(\bL)$.
\end{Thm}
%
%
%
\section{Conclusion}
\par
We have discussed methods towards an asymptotic theory of quantum Gibbs 
families in the ultra-cold regime. Some properties, such as the continuity 
of $\rho^*$, break down from finite temperatures to absolute zero while 
others, such as the Pythagorean and projection theorem, extend. 
In this article we have extended a representation of the irreducible correlation.
\par
Comparisons to models of quantum statistical physics will show how to make 
further developments. On the other hand basic mathematical questions are 
widely unexplored such a the continuity of the irreducible correlation of 
three qubits or the continuity of the maximum-entropy inference beyond the
case of two qutrit Hamiltonians. 
%
%

\begin{Ack}
This work was supported by the DFG project ``Quantum Statistics: 
Decision problems and entropic functionals on state spaces''. Thanks to 
B.\ Zeng for her helpful letter about the ideas in \cite{Chen2014} 
and to the referees for their valuable comments. 
\end{Ack}


%
%
\bibliographystyle{aipproc}   

\end{document}